\documentclass[useAMS,usenatbib]{mn2e}

\usepackage{amssymb}
\usepackage{graphicx}
\title[X-ray jets from Cir X-1]{A parsec scale X-ray extended structure from the X-ray binary Circinus X-1}
\author[P. Soleri et al.]{P. Soleri$^{1}$\thanks{E-mail:
p.soleri@uva.nl}, S. Heinz$^{2}$, R. Fender$^{3,1}$, R. Wijnands$^{1}$, V. Tudose$^{1,4,5}$,
D. Altamirano$^{1}$, \newauthor P.G. Jonker$^{6,7}$, M. van der Klis$^{1}$, L. Kuiper$^{6}$, C. Kaiser$^{3}$, P. Casella$^{1}$\\
$^{1}$Astronomical Institute `Anton Pannekoek', University of
Amsterdam, Kruislaan 403, 1098 SJ, Amsterdam, the Netherlands\\
$^{2}$Astronomy Department, University of Wisconsin-Madison, 475 N. Charter St. Madison, WI 53705, US\\
$^{3}$School of Physics and Astronomy, University of Southampton, Hampshire, SO17 1BJ, UK\\
$^{4}$Astronomical Institute of the Romanian Academy, Cutitul de Argint 5, RO-040557 Bucharest, Romania\\
$^{5}$Research Center for Atomic Physics and Astrophysics, Atomistilor 405, RO-077125 Bucharest, Romania\\
$^{6}$SRON, Netherlands Institute for Space Research, Sorbonnelaan 2, 3584 CA, Utrecht, the Netherlands\\
$^{7}$Harvard-Smithsonian Center for Astrophysics, Cambridge, MA 02138, US
}

\begin{document}

\date{Accepted 2008 October 13. Received 2008 September 22; in original form 2008 May 30}

\pagerange{\pageref{firstpage}--\pageref{lastpage}} \pubyear{2008}

\maketitle

\label{firstpage}

\begin{abstract} 
We present the results of the analysis of two {\it Chandra} observations of Circinus X-1 performed in 2007, for a
total exposure time of $\sim$50 ks. The source was observed with the High Resolution Camera during a long X-ray
low-flux state of the source. Cir X-1 is an accreting neutron-star binary system that exhibits ultra-relativistic
arcsec-scale radio jets and an extended arcmin-scale radio nebula. Furthermore, a recent paper has
shown an X-ray excess on arcmin-scale prominent on the side of the receding radio jet. In our images we clearly
detect X-ray structures both on the side of the receding and the approaching radio jet. The X-ray emission is
consistent with being from synchrotron origin.
Our detection is consistent with neutron-star binaries being as efficient as black-hole binaries in producing X-ray
outflows, despite their shallower gravitational potential.
\end{abstract}


\section{Introduction} \label{par:intro}
Cir X-1 is an exotic X-ray binary discovered by Margon et al. (1971) which shows flares
with a period of 16.55 days, observed first in the X-ray band (Kaluzienski et al. 1976) and then in the
infrared (Glass 1978; Glass 1994), radio (Haynes et al. 1978) and optical bands (Moneti 1992): this fact
is interpreted as enhanced accretion close to the periastron passage of a highly eccentric
binary orbit ($e \, \sim \, 0.8$, Murdin et al. 1980, Nicolson, Glass \& Feast 1980). 
Beyond variability at the 16.55 day orbital period the source shows dramatic evolution of its
X-ray luminosity, spectra and timing properties on timescales from milliseconds to decades
(Shirey et al. 1999, Parkinson et al. 2003).  The evidence that the system
harbours a neutron star comes from the detection of type-I X-ray bursts (Tennant et al.
1986a,b), confirmed by the recent detection of twin kHz QPOs in the X-ray power density spectra
(Boutloukos et al. 2006). Based on the properties of the type-I X-ray bursts, Jonker \&
Nelemans (2004) estimated a distance to the source of 7.8-10.5 kpc; according to a
different measure of the galactic absorption, Iaria et al. (2005) derived a lower distance
of 4.1 kpc (but see Jonker, Nelemans, Bassa 2007).

%We will adopt a distance D = 7.8 kpc throughout this paper.
% in the next para possible references to McCollough et al. 1999 and Abell & Margon 1979, for
% Cyg X-3 and SS 433 respectively
Cir X-1 is the most radio-loud neutron star X-ray binary (Whelan et al.
1977, Haynes et al. 1978; Cyg X-3 and SS433 are brighter in radio but their nature is unclear),
showing extended structures both at arcmin and arcsec scale.
The arcmin-scale structure has been extensively studied by Stewart et al (1993) and by Tudose
et al. (2006): the source shows two radio jets (south-east and north-west direction)
embedded in a large scale, diffuse radio nebula. There is general agreement that this
nebula is the result of the radio lobe inflated by the jets over several hundred-thousand years.
Arcmin-scale jets are curved which might be due to an interaction with
the interstellar medium (ISM) or a precessing jet. The arcsec-scale structure has been investigated for the
first time by Fender et al. (1998) and later studied by Fender et al. (2004), that reported a one-sided
highly-variable jet (for a detailed study of the arcsec-variability evolution over 10 years
see Tudose et al. 2008), the most
relativistic one detected so far within our Galaxy, with a bulk Lorentz factor $\geq$ 10.

The X-ray light curve of Cir X-1 is highly variable and characterized by bright flares and periods
of very low X-ray flux (Parkinson et al. 2003). On 2005 June 2nd Cir X-1 was observed with the High
Energy Transmission Gratings (HETGS) onboard {\it Chandra} for 50 ks during one of its long-term X-ray
low-flux states (an example of X-ray light curve with long periods of low flux is reported in Figure
\ref{fig:licu}). Analysing the HETGS observation, Heinz et al. (2007) found evidence for an arcmin-scale
X-ray structure prominent on the side of the receding radio jet. While for black hole candidates
extended X-ray jets have already been detected with {\it Chandra} and {\it XMM-Newton} in a number of sources (e.g. XTE
J1550-564, Corbel et al. 2002; 4U 1755-33, Angelini \& White 2003; H1743-322, Corbel et al. 2005),
this is the first detection of X-ray structures in a secure neutron star system, showing that neutron stars can be
as efficients as black holes in producing X-ray outflows.\\
Here we present recent {\it Chandra} observations of Cir X-1 where
we clearly detect an extended X-ray structure both on the side of the approaching and the receding
radio jet, confirming and extending the detection of Heinz et al. (2007).

% CIAO = Chandra Interactive Analysis of Observations
\section{Observation and data analysis} \label{par:analysis}
%########### FIGURE ##########################
\begin{figure}
\begin{tabular}{c}
\resizebox{8.4cm}{!}{\includegraphics{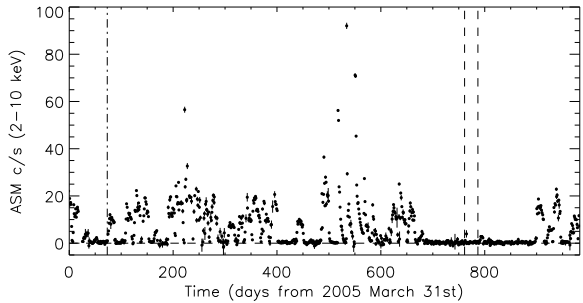}}
\end{tabular}
\caption{RXTE/ASM daily-average light curve from 2005 March 31 to 2007 November 29. The vertical
dashed lines correspond to two {\it Chandra}/HRC-I observations, the vertical dashed-dotted line corresponds
to the HETGS observation analysed in Heinz et al. (2007).}
\label{fig:licu}
\end{figure}
%#############################################
Cir X-1 was observed with the High Resolution Camera (HRC; Zombeck et al. 1995, Murray et al. 1998)
onboard the {\it Chandra} X-ray Observatory on 2007 April 21st (43 ks)
and on 2007 May 16th (7 ks), during an exceptionally long interval of very low X-ray flux that occurred from
$\sim$ 2007 February until $\sim$ 2007 August.
Figure \ref{fig:licu} shows the X-ray light curve of Cir X-1 taken with the All Sky Monitor (ASM) instrument
onboard the {\it Rossi X-ray Timing Explorer} satellite (RXTE) for the period 2005 March - 2007 November, where several
episodes of low-flux can be identified. Two vertical dashed lines correspond to our two HRC-I observations
(hereafter observation A and observation B, respectively).

\subsection{X-ray jets}
\begin{figure*}
\begin{tabular}{c}
\resizebox{12.0cm}{!}{\includegraphics{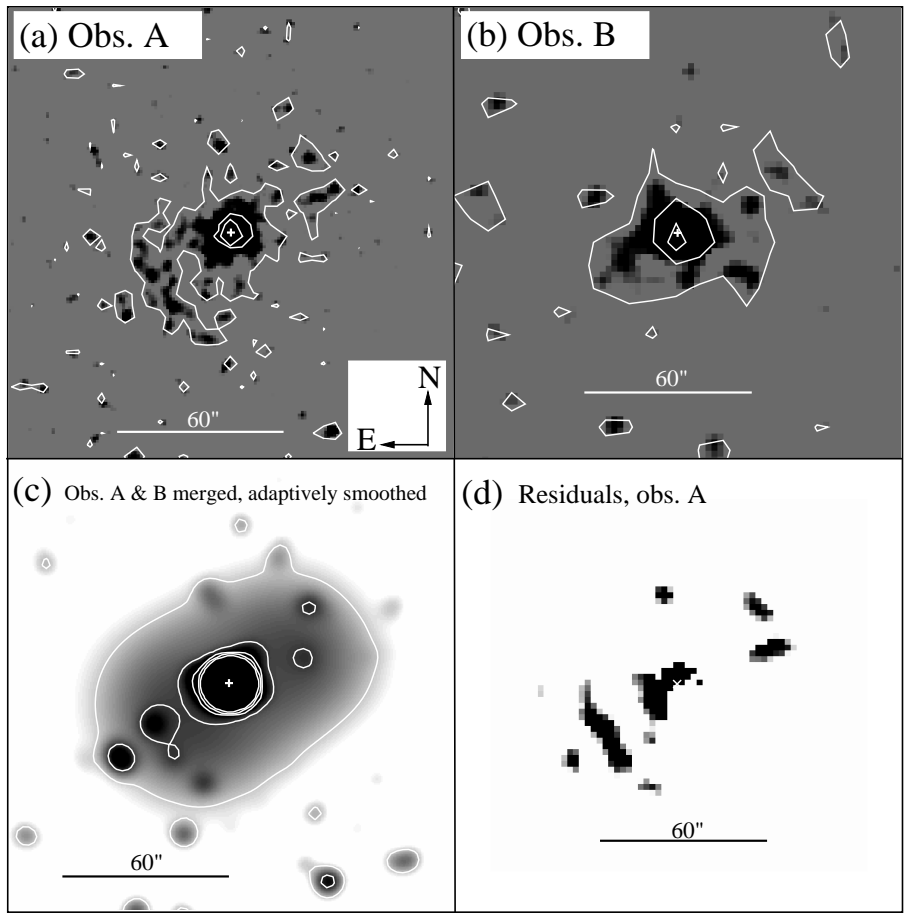}}
\end{tabular}
\caption{(a) {\it top left:} HRC-I image of obs. A. (b) {\it top right:} HRC-I image of obs. B. (c)
{\it bottom left:} image of the merger of obs. A and B, adaptively smoothed. (d) {\it bottom right:} image of the
residuals obtained after fitting the source image (obs. A) with a gaussian function and a constant using the PSF image as
the convolution kernel. In all the panels we marked the source location with a cross.}
\label{fig:smooth}
\end{figure*}

\begin{figure}
\begin{tabular}{c}
%\hspace{-0.7cm}
\resizebox{7.5cm}{!}{\includegraphics[angle=0]{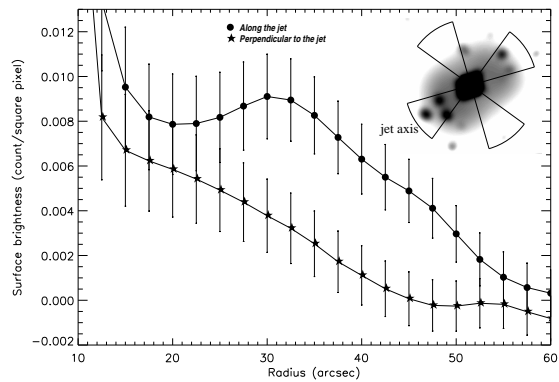}}
\end{tabular}
\caption{Radial surface-brightness profile extracted, for obs. A, across the jet (circles) and across a region
perpendicular to the jet (stars). The extraction regions are shown in the inset and correspond to the following PAs
(see definition in the text): X-ray excess 106$^{\circ}$-144$^{\circ}$ and 286$^{\circ}$-324$^{\circ}$;
axis perpendicular to the X-ray excess 16$^{\circ}$-54$^{\circ}$ and 196$^{\circ}$-234$^{\circ}$.}
\label{fig:profile}
\end{figure}

Images from observations A and B are reported in Figure \ref{fig:smooth}. The
images have been rebinned and smoothed using a gaussian kernel of three pixels in radius and countour lines
have been applied.
%########### FIGURE ##########################
%#############################################
A visual inspection of the images clearly reveals the presence of an extended X-ray structure
around the source up to $\sim$arcmin scale, visible in the South-East and in the North-West quadrant (i.e.
aligned with the South-East - North-West direction), consistent both with the results recently reported by Heinz et al. (2007)
and with the alignment of the arcmin-scale jet observed by Stewart et al. (1993) and by
Tudose et al. (2006); this issue will be discussed in \S\ref{par:comparison}.\\
The extended emission is detected both in obs. A and B and the two images (note the different
exposure in the two observations), at a visual inspection, are consistent: the X-ray emission around the
central source, in both of them, elongates along the same axis and furthermore, it presents similar
structures in the North-West quadrant and a similar spike-like structure in the northern quadrant (although there are
differences specially in the South-East quadrant, due to different sensitivity limits).\\
Since the two images are consistent, we can merge them to have an image with a higher equivalent exposure,
using the standard {\it Ciao 4.0} analysis tools (Fruscione et al. 2006). We adaptively smoothed the image
obtained from the merging with an average significance of $8.3 \sigma$ per smoothing length (the minimum smoothing
length is 6 pixels, the maximum one corresponds to the image size) and we applied contour levels.
The resulting image is plotted in Figure \ref{fig:smooth} (panel (c)), where a diffuse X-ray emission elongated
along the South-East - North-West direction is evident, extending up to $\sim$1 arcmin from the point source. Other
structures might look real from an analysis of this image (e.g a circular excess in the South-West quadrant) but they will not be
discussed since an inspection of the residuals obtained after fitting the point spread function (PSF)
suggests that they are noise features (see next part of this paragraph).\\
Figure \ref{fig:profile} shows two profile cuts (for obs. A), one extracted on a region aligned with the X-ray excess
and one along the perpendicular axis, showing an excess along the South-East - North-West axis and supporting the evidence that
there is an X-ray extended structure aligned with a privileged direction.\\
From Figure \ref{fig:smooth} (panels (a), (b) and (c)) and Figure \ref{fig:profile} the presence of an extended
X-ray structure around the central source is evident but to make our detection more robust, an analysis of the PSF is needed. 
We simulated a monochromatic PSF at 1 keV (from Heinz et al. 2007 we expect the X-ray excess emission to peak
at this energy) using {\it ChaRT} (Carter et al. 2003), considering the same number of counts as detected from the source
and we projected it in the detector
plane using the {\it Marx} ray tracing (Wise 1997). Since the PSF wings depend on the position on the detector, for the PSF
analysis we used only the obs. A: even if the image from the merging has a higher equivalent exposure,
the PSF extracted from that image might contain artefacts (the source position on the detector is not the same for
obs. A and B) that we want to avoid.
For the PSF analysis, the {\it Sherpa 3.4} tools (Freeman et al. 2001) have been used.\\
In Figure \ref{fig:smooth} (panel (d)) we show the image of the residuals obtained after fitting the
source image with a gaussian function and a constant using the PSF image as the convolution kernel. The presence of an
extended X-ray structure aligned with the South-East - North-West axis is evident, at a distance from the central source
between $\sim 25$" and $\sim 50$" (the lower limit is inferred by inspecting both the residuals image and the
radial profile image in Figure \ref{fig:profile}). The X-ray excess on the side of the receding radio-jet
(North-West quadrant, hereafter ``receding X-ray jet'') lies at position angle (PA, measured counterclockwise from due North from the
point source) intervals 286$^{\circ}$-295$^{\circ}$ and
307$^{\circ}$-324$^{\circ}$. A knot appears also at PA $\sim 10^{\circ}$ and its identification will be
discussed in the next sub-section. On the side of the approaching radio jet (South - East quadrant) we see one main
X-ray blob in a PA interval 88$^{\circ}$-152$^{\circ}$ (elongated along a North-East - South-West direction,
hereafter ``approaching X-ray jet''),
at a distance from the point source between 28'' and 38''. Other minor blobs appear in the residuals image: all of them are
located in the South - East quadrant and are consistent with being knots of the approaching X-ray jet.
%########### FIGURE ##########################
%#############################################

\subsection{Comparison with previously detected X-ray extended structure and the radio jet} \label{par:comparison}
\begin{figure}
\begin{tabular}{c}
\resizebox{7.0cm}{!}{\includegraphics{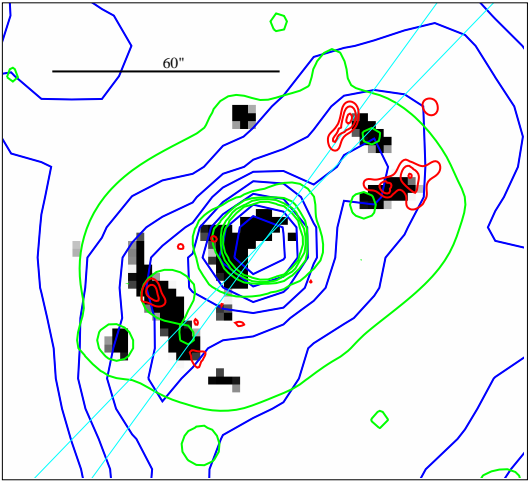}}
\end{tabular}
\caption{Radio-X-ray overlay. Grey scales: X-ray residuals image (Fig. \ref{fig:smooth} (d)); green contours:
adaptively smoothed image of obs. A \& B merged (Fig. \ref{fig:smooth} (c)); red contours: adaptively
smoothed, normalized, PSF subtracted image (from Heinz et al. 2007); dark blue contours: 1.4 GHz
surface brightness (adapted from Tudose et al. 2006, levels increase by $\sqrt{2}$ between contours, outermost
contour: 11.2 mJy/beam); light blue lines: estimated allowed range of PAs from high-resolution radio observations of
approaching radio-jet (Fender et al. 2004). The crossing of these two lines identify the source position.}
\label{fig:X_and_radio}
\end{figure}

The radio nebula and the relativistic radio-jets of Cir X-1 have been investigated during multiple epochs of radio
observations, both on arcmin and arcsec scale (Stewart et al. 1993, Fender et al. 1998,
Fender et al. 2004, Tudose et al. 2006, Tudose et al. 2008). From those papers, we estimate the corresponding PAs for
the arcmin-scale jets: $303^{\circ} \pm
9^{\circ}$ and $310^{\circ} \pm 15^{\circ}$ for the jets in Stewart et al. (1993) and Tudose et
al. (2006) respectively and $320^{\circ} \pm 4^{\circ}$ for the ultrarelativistic jets in Fender et al. (2004). 
All these angles are consistent with the PAs of the X-ray jets in our HRC-I observations and fall between
the two X-ray filaments detected by Heinz et al. (2007) in the HETGS image.\\
Figure \ref{fig:X_and_radio} shows an overlay of the residuals image (the same as Figure \ref{fig:smooth} panel (d),
grey scale), the X-ray contours from Figure \ref{fig:smooth} panel (c) and the jet emission from
Figure 1 of Heinz et al. (2007). Also shown are the radio contours from Figure 3 of Tudose et al.
(2006) and the limits on the PA for the arcsec-scale jet (Fender et al. 2004). The X-ray jets detected
in the HRC-I images are broadly consistent with the X-ray excess of Heinz et al. (2007). Besides the X-ray knot that
we detected
at PA 10$^{\circ}$, the consistency between the X-ray and the radio jets is clear on both sides
(approaching and receding): what we see in the HRC-I images appears as the X-ray counterpart of the radio
jets from Cir X-1.
%########### FIGURE ##########################
%#############################################

\section{Discussion and conclusions}
The analysis of our two HRC-I observations of Cir X-1 clearly showed X-ray jets both on the side of the receding and the
approaching radio jet, consistent with being the X-ray counterpart of the arcmin-scale radio jets
(Tudose et al. 2006). Heinz et al. (2007)  proposed two
possible alternative explanations for the origin of the X-ray excess: synchrotron emission and thermal
bremsstrahlung. We now investigate whether these two mechanisms can still explain the X-ray jets as observed by HRC-I.

{\bf Synchrotron emission:} HRC-I does not have the energy resolution necessary to allow for spectral fitting,
instead we calculate the jet flux with {\it webPIMMS}\footnote{http://heasarc.nasa.gov/Tools/w3pimms.html}
using, as input spectrum, the absorbed power law fitted in Heinz et al. (2007)
($\Gamma = 3.0^{+2.6}_{-1.5}$, $N_{H} = 5.9 \times 10^{22}$ cm$^{-2}$)
and the average count rate extracted, for obs. A, in the region corresponding to the main knot of the approaching jet, using 
{\it Ciao 4.0}. This gives
an un-absorbed flux $F_{2-10 keV} = 7.4 \times 10^{-13}$ erg cm$^{-2}$ s$^{-1}$, corresponding to an X-ray luminosity 
$L_{2-10 keV} = 5.4 \times 10^{33}$ erg s$^{-1} D^2_{7.8}$ (throughout the paper we adopt a distance from the source
$D = 7.8$ kpc, the same adopted in Heinz et al. 2007, to make comparisons easier). 
The used photon index $\Gamma$ is broadly consistent with the emission being of synchrotron origin
($\Gamma_{syn} \sim 1.5$); assuming a source volume $V$, a specific luminosity $L_{\nu}$ and a
spectrum of the form $L_{\nu} \propto \nu^{\alpha}$, ($\alpha = 1 - \Gamma$), we can follow Longair (1994) and Fender
(2006) to estimate the minimum energy associated with the source. The morphology of the jet suggests that we are
observing the surface of a conical volume where the emission takes place. Considering the jet length to be roughly $l_{jet}
\approx 1.6$ pc $D_{7.8}/ sin\, i$ (where $i$ is the angle with the line of sight) and a half-opening angle $\beta
\approx 19.3 ^{\circ}$ the volume of the emitting cone is $V_{jet} = 9.7 \times 10^{55} $cm$^3 D^{3}_{7.8}
/ sin\, i$, assuming equal emitting volumes both on the receding and on the approaching side. Under these assumptions
the minimum jet energy is $E_{min} \gtrsim 6 \times 10^{44}$ erg. Again following Longair (1994) and Fender (2006), 
we can calculate the magnetic field $B_{min}$ associated to $E_{min}$, the Lorentz factor $\gamma$
of the energetic electrons emitting by synchrotron and their gyro-radius $r_g$: $B_{min} = 8.2 \mu$G, $\gamma =
2.3 \times 10^{8}$ and $r_g =  1.6\times 10^{-2}$ pc. The energy of
these electrons is $E_{e} = 1.2 \times 10^{14}$ eV and their energy loss rate is $R \sim 2300$ eV/s: considering that we
have no evidence for re-acceleration taking place, the lifetime $t_{syn}$ of these electrons is $t_{syn} \approx
1600$ yr. Following Tudose et al. (2006), we assume that the jets are injected mainly during the
flare states (duty cycle $\sim$6\%): the resulting minimum jet power is $W_{syn} \gtrsim 2 \times 10^{35}$ erg/s.
 
{\bf Thermal bremsstrahlung:} in this case the X-rays would originate from the shock driven into the ISM by the
propagation of the jets. Here we use a model developed for
extragalactic jet sources by Castor et al. (1975), Kaiser \& Alexander (1997) and Heinz et al. (1998). We assume the
temperature of the thermal gas to be $k_{B}T_{shock} = 2.2$ keV ($k_{B}T_{shock} = 2.2^{+7.0}_{-1.1}$ keV and
$N_{H} = 5.4 \times 10^{22}$ cm$^{-2}$ are obtained by
Heinz et al. 2007 fitting the jet spectrum with an absorbed thermal model), the electron density to be
$n_{e} \approx 10$ cm$^{-3}$ (as used in Heinz et al. 2007; since the temperature $T_{shock} = 25.5$ MK we assume
an ionization fraction x=1: all the hydrogen is ionized) and the length of the shock region (considering both
receding and approaching side) $L_{shock} \approx 2.27$~pc. Such length and the used density give an emitting gas mass of
0.02-0.4 $M_{\odot} D^{5/2}_{7.8}/ sin\, i$: uncertainties are due to different
possible measures of the thickness of the shock region.  
Following Kaiser \& Alexander (1999) and balancing the interior pressure exerted by the jets and the ram pressure
of the shocked ISM, the jet lifetime is $t_{th} = \frac{3}{5} \frac{L}{v_{sp}} \approx 1700$ yr ($v_{sp}$ is
the velocity of the shock-compressed particles, obtained from the shock temperature $k_{B}T_{shock}=2.2$ keV) and the
jet power is $W_{th} \approx 6 \times 10^{37}$ erg/s.

{\bf Discussion:} In the synchrotron case the jet lifetime $t_{syn}$ is smaller than the time expected for the jet to
inflate the large-scale radio lobe ($10^4 - 10^5$ yr, Tudose et al. 2006) and this suggests that the X-ray emission could
come from the jet itself rather than from the inflated radio nebula. Furthermore, the
gyro-radius of the electrons responsible for the X-rays is smaller than the size of the jet ($r_g < l_{jet}$) and this implies
that those electrons can be confined in the jet region.
Our jet power value is consistent with the estimate of Tudose et al. (2006) ($W_{jet,Tudose} \sim 10^{35}$ erg/s, which however
could be up to two order of magnitude higher) and one order of magnitude smaller than the jet power calculated by Heinz et al.
(2007) ($W_{syn,Heinz} \gtrsim 5 \times
10^{36}$ erg/s). The magnetic field associated to the emitting electrons is again consistent with the value reported in Tudose
et al. (2006) ($B_{min,Tudose} = 6.3 \mu$G), making $W_{syn}$ a robust estimate of the jet power, sufficient to inflate the
radio nebula. $W_{syn }$ is $\gtrsim 0.1$\% of the Eddington luminosity ($L_{Edd} = 1.8 \times 10^{38}$ erg/s)
for a 1.4$M_{\odot}$ neutron star: Cir X-1 is only slightly super-Eddington and this would imply a jet-production
efficiency $\eta \gtrsim \, $0.01\%, consistent with what obtained by Heinz et al. 2007 ($\eta \gtrsim 0.5$ \%).\\
In the thermal bremsstrahlung case we calculated a jet power $W_{th} \approx 6 \times 10^{37}$ erg/s, two orders of magnitude
larger than $W_{syn}$ and possibly not consistent with the values estimated by Heinz et al. (2007)
($W_{th,Heinz} = 5 \times 10^{36}$ erg/s). Furthermore, $W_{th}$ is a significant fraction ($\sim 33$\%) of the Eddington
luminosity ($L_{Edd} = 1.8 \times 10^{38}$ erg/s) for a 1.4$M_{\odot}$ neutron star
and this would be an extremely high jet-production efficiency $\eta \approx \, $3\%, even for an accreting black hole.
Taking an Eddington-limit mass accretion rate $\dot{M} = 10^{18}$ g s$^{-1}$ and even considering that all the accreted mass is
ejected in an outflow, the time required to inflate the jets would be $\sim$2.7 Myr, much bigger than
$t_{th}\approx 1700$ yr, suggesting that the emission can not come from the jet itself.
Therefore thermal bremsstrahlung appears as an unlikely emission mechanism (especially compared to the synchrotron
case) for the X-ray jets. 

\section*{Acknowledgments}
The authors thank Harvey Tannanbaum for scheduling the DDT observations and Alessandro Patruno
and David Russell for very useful comments and discussion. PS would also like to thank Nanda Rea, Simone Migliari and
the {\it Chandra} Help Desk for very useful suggestions on the {\it Chandra} data analysis.
PS and PGJ also acknowledge support from NWO.
% ####### Bibliography ########################################################

% ######## End bibliography ###################################################
\label{lastpage}

\end{document}